\documentclass[aps,twocolumn,pra,superscriptaddress,showpacs,tightenlines]{revtex4}
\usepackage{amssymb}
\usepackage{amsmath}
\usepackage{graphicx}
\usepackage{epsfig}
\usepackage{subfigure}
\usepackage{amsfonts}

\input{tcilatex}
\begin{document}

\title{Effective Hamiltonian approach to the Kerr nonlinearity in an
optomechanical system}
\author{Z. R. Gong}
\affiliation{Advanced Science Institute, The Institute of Physical and Chemical Research
(RIKEN), Wako-shi, Saitama 351-0198, Japan}
\affiliation{Institute of Theoretical Physics, The Chinese Academy of Sciences, Beijing,
100080, China}
\author{H. Ian}
\affiliation{Advanced Science Institute, The Institute of Physical and Chemical Research
(RIKEN), Wako-shi, Saitama 351-0198, Japan}
\affiliation{Institute of Theoretical Physics, The Chinese Academy of Sciences, Beijing,
100080, China}
\author{Yu-xi Liu}
\affiliation{Advanced Science Institute, The Institute of Physical and Chemical Research
(RIKEN), Wako-shi, Saitama 351-0198, Japan}
\affiliation{Institute of Microelectronics, Tsinghua University, Beijing 100084, China}
\author{C. P. Sun}
\affiliation{Advanced Science Institute, The Institute of Physical and Chemical Research
(RIKEN), Wako-shi, Saitama 351-0198, Japan}
\affiliation{Institute of Theoretical Physics, The Chinese Academy of Sciences, Beijing,
100080, China}
\author{Franco Nori}
\affiliation{Advanced Science Institute, The Institute of Physical and Chemical Research
(RIKEN), Wako-shi, Saitama 351-0198, Japan}
\affiliation{Center for Theoretical Physics, Physics Department, The University of
Michigan, Ann Arbor, MI 48109-1040, USA.}
\date{\today}

\begin{abstract}
Using the Born-Oppenheimer approximation, we derive an effective Hamiltonian
for an optomechanical system that leads to a nonlinear Kerr effect in the
system's vacuum. The oscillating mirror at one edge of the optomechanical
system induces a squeezing effect in the intensity spectrum of the cavity
field. A near-resonant laser field is applied at the other edge to drive the
cavity field, in order to enhance the Kerr effect. We also propose a
quantum-nondemolition-measurement setup to monitor a system with two
cavities separated by a common oscillating mirror, based on our effective
Hamiltonian approach.
\end{abstract}

\maketitle

%\global\columnwidth20.5pc \global\hsize\columnwidth\global\linewidth%
%\columnwidth
%\global\displaywidth\columnwidth

\section{\label{sec:introduction}INTRODUCTION}

Recently, experimental and theoretical investigations have been carried out
to demonstrate the coherent optomechanical coupling between a quantized
cavity field and the mechanical motion of a mirror located at one end of an
optical cavity (e.g., see Refs.~\cite%
{Meystre,optomechanical-1,optomechanical-2,optomechanical-3}). Many efforts
have been made to propose new devices and explore the quantum-classical
transition (e.g., see Refs.~\cite{h-1,h-2,m-1,m-2}) based on such
optomechanical systems. It was also discovered that such systems possess
nonlinear optical properties. Our study here focuses on another aspect of
optomechanical coupling that can realize the Kerr nonlinear optical effect
in a vacuum, rather than in a conventional dispersive medium~\cite%
{Kerr-2,Kerr-3}.

The Kerr effect usually appears in nonlinear dispersive media~\cite%
{nonlinear}, due to the third-order matter-light interaction. Instead of
resulting from higher-order light-matter interactions, we now realize
\textit{the nonlinear Kerr effect} from the radiation pressure on an
oscillating mirror. This oscillating mirror, located at one end of a Fabry-P%
\'{e}rot (FP) cavity, is driven by an input laser field at the other end.
The oscillating mirror is modeled as a quantum mechanical resonator, whose
tiny oscillations are controlled by the radiation pressure of the cavity
field. Indeed, it has been pointed out~\cite{Meystre-1,Mancini-1} that the
vacuum cavity with a movable mirror might mimic a Kerr medium when the
cavity field is driven by a coherent light field.

In this article, we use the Born-Oppenheimer (BO) approximation to derive a
Kerr-medium-like Hamiltonian, which shows the underlying nonlinear mechanism
more clearly than other (equivalent) approaches.

We shall first point out first that, in the conventional BO approximation
for a molecule the (slower) nuclear variables are adiabatically separated
from the (faster) electronic variables. The stability of the molecular
configuration requires the effective potential to have a minimum value.
However, the generalized BO theory~\cite{newBO} for spin-orbit systems or
cavity QED systems~\cite{QED} does \textit{not} have this requirement, and
the effective force could be either attractive or repulsive.

\section{\label{sec:two}INDUCED KERR NONLINEARITY AND BORN-OPPENHEIMER
APPROXIMATION}

As shown in Fig.~\ref{fig:fig1}(a), we study a Fabry-P\'{e}rot (FP) cavity
with an oscillating mirror at one end, acting as a quantum-mechanical
harmonic oscillator. The cavity is driven by a laser field with frequency $%
\omega _{\mathrm{D}}$. We first study an ideal case, without considering the
losses of both the cavity field and the oscillating mirror. The total
Hamiltonian
\begin{equation}
H=H_{\mathrm{P}}+H_{\mathrm{M}}+H_{\mathrm{I}}  \label{2-1}
\end{equation}%
contains three parts as
\begin{subequations}
\begin{align}
H_{\mathrm{P}}& =\hbar \omega _{0}a^{\dag }a-\hbar ga^{\dag }ax,
\label{2-2-2} \\
H_{\mathrm{M}}& =\frac{p^{2}}{2m}+\frac{1}{2}m\Omega ^{2}x^{2},
\label{2-2-1} \\
H_{\mathrm{I}}& =i\hbar \lambda \left( a^{\dag }e^{-i\omega _{\mathrm{D}%
}t}-ae^{i\omega _{\mathrm{D}}t}\right) .  \label{2-2-3}
\end{align}%
The cavity field with frequency $\omega _{0}$ is described by bosonic
operators $a$ and $a^{\dag }$. The symbols $m$, $\Omega $, $x$, and $p$
denote, respectively, the mass, frequency, displacement, and momentum of the
oscillating mirror (hereafter, we just call it \textquotedblleft the
mirror"). The coupling strength $\lambda $ between the cavity field and the
driving laser field is related to the input laser power $P$ and the decay
rate $\kappa $ of the cavity field via the relation $\left\vert \lambda
\right\vert =\sqrt{2P\kappa /\hbar \omega _{\mathrm{D}}}$. The interaction
constant $g=\omega _{0}/L$ between the cavity field and the mirror stems
from a very small change $x$ of the FP cavity length $L$.

%%%%%%%%%%%%%%%%%%%%%%%%%%%%%%%%
\begin{figure}[ptb]
\begin{centering}
\includegraphics[bb=49bp 612bp 539bp 760bp,width=3.2in]{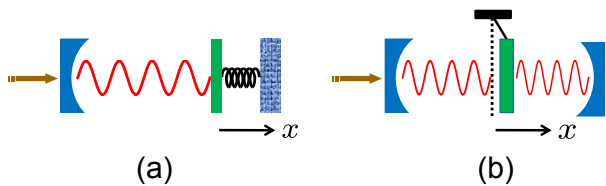}
\end{centering}
\caption{(Color online) Schematic diagram of (a) a Fabry-P\'{e}rot
cavity with an oscillating mirror (green) in one end, and (b) two
Fabry-P\'{e}rot cavities sharing a common oscillating mirror. The
mirrors in (a) and (b) are modeled as quantum-mechanical harmonic
oscillators. The red sinusoidal curves inside the cavities and the
spring schematically represent the cavity field and the harmonic
oscillator, respectively. The cavity is driven by an input laser
field shown as a (brown) arrow on the left. The two oscillating
mirrors have very small displacements, $x,$ around their equilibrium
positions.} \label{fig:fig1}
\end{figure}
%%%%%%%%%%%%%%%%%%%%%%%%%%%%%%%%%%%

Usually, the characteristic frequency of the cavity field is about~$10^{14}$
Hz, which is much higher than the nanomechanical resonator frequency~$10^{9}$
Hz achieved by current experiments. The cavity field would have \textit{no}
nonlinear effect under the BO approximation when there is no driving field;
nonlinear effects appear when a classical driving field is applied to the
cavity. Let us consider the case when the driving field frequency $\omega _{%
\mathrm{D}}$ is close to the cavity field frequency $\omega _{0}$. We also
use the \textquotedblleft rotating frame of reference\textquotedblright\
defined by a unitary transformation $W(t)=\exp (-i\omega _{\mathrm{D}%
}a^{\dag }a\,t)$, which is very similar to the NMR experiments used to
demonstrate the Berry phase~\cite{berry-0}. In the rotating frame of
reference, the effective form $H^{\mathrm{R}}=W^{\dag }(t)HW(t)-iW^{\dag
}(t)(\partial W(t)/\partial t)$ of the Hamiltonian $H$ in Eq.~(\ref{2-1})
reads
\end{subequations}
\begin{equation}
H^{\mathrm{R}}=H_{\mathrm{C}}+H_{\mathrm{MR}}^{\mathrm{R}},  \label{2-3}
\end{equation}%
with the effective Hamiltonians
\begin{equation}
H_{\mathrm{C}}=\hbar \Delta a^{\dag }a+i\hbar \lambda (a^{\dag }-a),
\label{2-4}
\end{equation}%
and
\begin{equation}
H_{\mathrm{MR}}^{\mathrm{R}}=\frac{p^{2}}{2m}+\frac{1}{2}m\Omega
^{2}x^{2}-\hbar ga^{\dag }a\,x\,.  \label{2-5}
\end{equation}%
Here, the detuning $\Delta =\omega _{0}-\omega _{\mathrm{D}}$ is the
effective frequency of the cavity field in the new frame. Clearly, $\Delta $
can be controlled by tuning the frequency $\omega _{\mathrm{D}}$ of the
driving field. Therefore, the effective frequency $\Delta $ of the cavity
field can be tuned to be much smaller than that of the mechanical resonator.
Under such condition, the mechanical resonator can be treated as the fast
variable and the BO approximation can be employed.

We first study the Hamiltonian~(\ref{2-5}) of the fast variables $x$ and $p$
of the mirror (in the rotating frame) by taking the \textquotedblleft slow
variables\textquotedblright\ $a$ and $a^{\dag }$ of the cavity field as
constants (in the rotating frame). Then the Hamiltonian~(\ref{2-5}) can be
rewritten as
\begin{align}
H_{\mathrm{MR}}^{\mathrm{R}}& =\hbar \Omega \left( A^{\dagger }A+\frac{1}{2}%
\right) -\frac{\hbar ^{2}g^{2}}{2m\Omega ^{2}}(a^{\dag }a)^{2},  \label{2-6}
\end{align}%
where the creation operator $A^{\dagger }$ of the cavity field is defined by
\begin{equation}
A^{\dag }=\sqrt{\frac{m\Omega }{2\hbar }}\left( x-\frac{ip}{m\Omega }\right)
+\alpha  \label{2-7}
\end{equation}%
with $\alpha =-\sqrt{2\hbar g^{2}N^{2}/m\Omega ^{3}}$. Eq.~(\ref{2-6}) shows
that the mirror variables are shifted by the amount $\alpha $ due to its
interaction with the cavity field. It is clear that the ground state of the
effective Hamiltonian in Eq.~(\ref{2-6}) can be obtained via the ground
state of the Hamiltonian $H_{\mathrm{M}}$ in Eq.~(\ref{2-2-1}) for a
harmonic oscillator with displacement operator
\begin{equation}
D(\alpha )=\exp (i\alpha A^{\dag }-i\alpha ^{\ast }A).  \label{2-8}
\end{equation}%
The eigenvectors and the eigenvalues of the mirror, corresponding to the
Hamiltonian~(\ref{2-6}), are, respectively,
\begin{equation}
\left\vert n\right\rangle =\frac{1}{\sqrt{n!}}(A^{\dag })^{n}D(\alpha
)\left\vert 0\right\rangle \equiv \left\vert n(a^{\dag }a)\right\rangle
\label{2-9}
\end{equation}%
and
\begin{equation}
V_{n}(a^{\dag }a)=\hbar \Omega \left( n+\frac{1}{2}\right) -\frac{\hbar
^{2}g^{2}}{2m\Omega ^{2}}(a^{\dag }a)^{2}.  \label{2-10}
\end{equation}%
Eqs.~(\ref{2-9}) and (\ref{2-10}) show that $\left\vert n\right\rangle $ and
$V_{n}(a^{\dag }a)$ are functions of the slow variables of the cavity field.
Eq.~(\ref{2-9}) also shows that the ground state of the fast variables ($x$
and $p$) of the mirror is a coherent state $D(\alpha )|0\rangle $, resulting
from radiation pressure.

According to the lowest-order generalized BO approximation, the total
eigenfunction $\left\vert \Phi \right\rangle $ of the coupled system of the
cavity field and the mirror can be factorized as $\left\vert \Phi
\right\rangle =\left\vert \phi _{n}(\alpha )\right\rangle \left\vert
n(a^{\dag }a)\right\rangle $, where $\left\vert \phi _{n}(\alpha
)\right\rangle $ satisfies the Schr\"{o}dinger equation with the effective
Hamiltonian
\begin{align}
H_{\mathrm{ph}}^{\mathrm{R}}& =\hbar \Delta a^{\dag }a+i\hbar \lambda
(a^{\dag }-a)+V_{n}(N)  \notag \\
& =\hbar \Delta a^{\dag }a-\hbar \chi (a^{\dag }a)^{2}+i\hbar \lambda
(a^{\dag }-a)+\mathrm{const}.  \label{2-11}
\end{align}%
The BO adiabatic separation provides an effective potential $V_{n}(a^{\dag
}a)$ for the \textquotedblleft slow\textquotedblright\ motion of the cavity
field. This potential contains a typical
\begin{equation}
\mathrm{\ Kerr\,\,nonlinear\,\,term}=\hbar \chi (a^{\dag }a)^{2},
\label{2-12}
\end{equation}%
where the parameter
\begin{equation}
\chi =\hbar \frac{g^{2}}{2m\Omega ^{2}}  \label{2-13}
\end{equation}%
plays the role of \textit{the phenomenological third-order susceptibility}
as in usual Kerr media.

\section{\label{sec:three} THE VALIDITY OF BO APPROXIMATION}

%%%%%%%%%%%%%%%%%%%%%%%%%%%%%%%%%%%%%%%%%%%%%%%%%%%%
\begin{figure}[ptb]
\begin{centering}
\includegraphics[bb=54bp 430bp 544bp 767bp,width=3.2in]{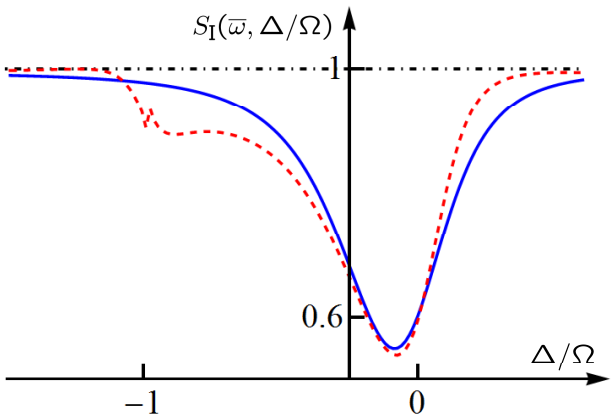}
\end{centering}
\caption{(Color online) Intensity spectrum $S_{\mathrm{I}}(\overline{\protect%
\omega},\Delta/\Omega)$ versus normalized detuning $\Delta/\Omega$ for the
value $\protect\omega=\overline{\protect\omega}$. Analytical results for the
full Hamiltonian are shown with a red dashed curve, while the BO
approximation results are shown with a blue solid curve. Note that squeezing
occurs when $S_{\mathrm{I}}(\overline{\protect\omega},\Delta/\Omega)<1$, and
the baseline $S_{\mathrm{I}}(\overline{\protect\omega},\Delta/\Omega)=1$ is
shown with a horizontal black dot-dashed line.}
\label{fig:fig2}
\end{figure}

%%%%%%%%%%%%% end of figure %%%%%%%%%%%%%%%%%%%%%%%%

We now verify the validity of the generalized BO approximation applied to
the optomechanical system through the squeezing effect, which is induced by
the Kerr interaction. This squeezing effect can be demonstrated by the
output intensity spectrum $S_{\mathrm{I}}(\omega ,\Delta )$. Following the
definition of the $S_{\mathrm{I}}(\omega ,\Delta )$ in Ref.~\cite{Mancini-1}
and the linearization technique of the Langevin equations governed by the
Hamiltonian in Eq.~(\ref{2-11}), we obtain the intensity spectrum under the
BO approximation
\begin{equation}
S_{\mathrm{I}}(\omega ,\Delta )=\left\vert 1-2\kappa \frac{A_{-}(\omega
,\Delta )+iB(\omega ,\Delta )\exp {i2\theta (\Delta )}}{D(\omega ,\Delta )}%
\right\vert ^{2},  \label{3-1}
\end{equation}%
where
\begin{subequations}
\begin{align}
A_{\pm }(\omega ,\Delta )& =-i\omega +i\Delta ^{\prime }\mp 4i\left\vert
\alpha _{\mathrm{s}}(\Delta )\right\vert ^{2}\chi +\kappa ,  \label{3-2-1} \\
B(\omega ,\Delta )& =2\alpha _{\mathrm{s}}(\Delta )^{2}\chi ,  \label{3-2-2}
\\
D(\omega ,\Delta )& =A_{+}(\omega )A_{-}(\omega )-\left\vert B(\omega
,\Delta )\right\vert ^{2},  \label{3-2-4} \\
\tan \frac{\theta (\Delta )}{2}& =\frac{\Delta ^{\prime }-2\chi \left\vert
\alpha _{\mathrm{s}}(\Delta )\right\vert ^{2}}{\kappa },  \label{3-2-5}
\end{align}%
with renormalized detuning $\Delta ^{\prime }=\Delta -\chi $ and the
steady-state value $\alpha _{\mathrm{s}}(\Delta )$ of the amplitude of the
cavity field. Here, the cavity loss $\kappa $ has been taken into account.
The BO approximation requires that the photons in the cavity can survive a
sufficiently long time, which is equivalent to the condition $\kappa \ll
\omega _{0}.$

If there is no Kerr interaction induced by the radiation pressure ($\chi =0$%
), $S_{\mathrm{I}}(\omega,\Delta)$ can be simplified to $S_{\mathrm{I}%
}(\omega,\Delta)=1$. When the Kerr interaction is induced ($\chi \neq 0$),
the intensity spectrum can be less than $1$, which displays a squeezing
effect~\cite{xuedong1,fxue07} of the cavity field. Here, $B(\omega,\Delta)$
plays an important role in the reduction of the output intensity
fluctuation. Hence, the squeezing effect could be observed experimentally by
measuring the intensity spectrum of the output laser.

Fig.~\ref{fig:fig2} shows the squeezing of the cavity field in the intensity
spectrum with a blue solid curve, where the maximum squeezing occurs in the
vicinity of $\left\vert \Delta /\Omega \right\vert \sim 0$. We consider that
the oscillating mirror has mass $m=100$ ng, frequency $\Omega /\kappa =40\pi
$, and a normalized damping rate $\gamma /\kappa =0.06$. Fig.~\ref{fig:fig2}
is plotted for the driving field frequency $\omega _{\mathrm{D}}/2\pi =282$
THz, the optical cavity length $L=10^{-2}$ m, finesse $\mathcal{F}=1.9\times
10^{5}$, and decay rate $\kappa \simeq 5\times 10^{5}$ s$^{-1}$; a driving
laser wavelength $\lambda =1064 $ nm, normalized frequency $\omega _{\mathrm{%
D}}/\kappa =3.54\times 10^{7}$ and power $P=500$ $\mu W$. The temperature $T$
of the cavity field and the mechanical resonator is assumed to be zero, as
in Ref.~\cite{parameter}. Then at a particular frequency $\overline{\omega }$
(chosen at the position where $S_{\mathrm{I}}(\omega,\Delta)$ is minimum),
the intensity spectrum has the squeezing effect shown in Fig.~\ref{fig:fig2}.

Now, we also study the intensity spectrum without the BO approximation.
Starting from the total Hamiltonian in Eq.~(\ref{2-3}), a full derivation
shows that the intensity spectrum has a similar form as in Eqs.~(15) except
that the renormalized detuning $\Delta ^{\prime }$ and the Kerr interaction
strength $\chi $ are replaced, respectively, by $\Delta -gx_{\mathrm{s}}$
and $\chi \Omega ^{2}\zeta (\omega ).$ Here, $x_{\mathrm{s}}=\chi \left\vert
a_{\mathrm{s}}\right\vert ^{2}$ is the steady-state value of the oscillation
amplitude of the mirror and
\end{subequations}
\begin{equation}
\zeta (\omega )=\frac{1}{\Omega ^{2}-\omega ^{2}-i\gamma \omega }
\label{3-3}
\end{equation}%
is the mechanical susceptibility of the oscillating mirror. The maximum
squeezing effect in the intensity spectrum, plotted by the red dashed curve
in Fig.~\ref{fig:fig2}, can also be observed in the vicinity of $\left\vert
\Delta /\Omega \right\vert \sim 0$. We use the same parameters for
calculating the intensity spectrum under the BO approximation. Therefore,
the BO approximation is valid when the frequency of the oscillating mirror $%
\Omega $ is much larger than the frequency of the light field $\Delta .$

The effect of the BO approximation can also be understood via $\zeta (\omega
)$. If the mirror frequency $\Omega $ is much larger than the detuning $%
\Delta $ of the cavity field in the rotating frame of reference, $\zeta
(\omega )$ plays an important role in the vicinity of $\omega \approx \Delta
$. When the BO approximation is valid under the condition $\Omega \gg \Delta
$, and when the macroscopic displacement $x_{\mathrm{s}}$ is extremely
small, the mechanical susceptibility is approximately equal to
\begin{equation}
\zeta (\omega )\approx 1/\Omega ^{2}  \label{3-4}
\end{equation}%
$,$ which leads the intensity spectrum to have the same form as in Eq.~(\ref%
{3-1}).

\section{\label{sec:fours}QUANTUM NONDEMOLITION MEASUREMENT WITH TWO-MODE
INDUCED KERR EFFECT}

Within the BO approximation, the cavity field inside the flexible Fabry-P%
\'{e}rot cavity, driven by an input laser, exhibits a Kerr-like nonlinear
property. It is not difficult to generate a two-mode induced Kerr
interaction, which is useful for quantum nondemolition (QND) measurements~%
\cite{measure-1}.

As shown in Fig.~\ref{fig:fig1}(b), we consider two FP cavities,
referred to below as the left and the right cavities with subindexes
$L$ and $R$, sharing one common oscillating mirror. This mirror is
assumed to oscillate with a very small displacement $x$ around its
equilibrium position. Thus two cavity fields, with frequencies
$\omega _{\mathrm{L}}$ and $\omega _{\mathrm{R}}$ in the laboratory
reference frame, indirectly interact with each other via this
oscillating mirror.

In the rotating frame of reference, using the BO approximation discussed
above, we can derive the induced effective interaction between the two
cavity fields
\begin{equation}
H_{\mathrm{eff}}(n_{\mathrm{L}},n_{\mathrm{R}})=-\hbar (\chi _{\mathrm{L}}n_{%
\mathrm{L}}^{2}+\chi _{\mathrm{R}}n_{\mathrm{R}}^{2}+2\sqrt{\chi _{\mathrm{L}%
}\chi _{\mathrm{R}}}n_{\mathrm{L}}n_{\mathrm{R}}),  \label{4-1}
\end{equation}%
where the
\begin{equation}
\chi _{i}=\hbar g_{i}^{2}/(2m\Omega ^{2})  \label{4-1-1}
\end{equation}
and $n_{i}$ with $i=\mathrm{L},\,\,\mathrm{R}$ denote the photon number
operator of the left and the right cavity fields. To perform a QND
measurement, the self-modulation term $\chi _{i}n_{i}^{2}$ of the probe
field can be ignored~\cite{measure-1}. Therefore, the system Hamiltonian for
the QND measurement can be written as
\begin{equation}
H_{\mathrm{QND}}=\hbar \Delta _{\mathrm{L}}a_{\mathrm{L}}^{\dag }a_{\mathrm{L%
}}+\hbar \Delta _{\mathrm{R}}a_{\mathrm{R}}^{\dag }a_{\mathrm{R}}+2\hbar
\chi a_{\mathrm{L}}^{\dag }a_{\mathrm{L}}a_{\mathrm{R}}^{\dag }a_{\mathrm{R}%
},  \label{4-2}
\end{equation}%
where $\Delta _{\mathrm{L}}=\omega _{\mathrm{L}}-\omega _{\mathrm{D}},$ and $%
\Delta _{\mathrm{R}}=\omega _{\mathrm{R}}-\omega _{\mathrm{D}}$ are the
detunings of the left and right cavity fields, relevant to the frequency $%
\omega _{\mathrm{D}}$ of the driving field.

The Kerr interaction in Eq.~(\ref{4-2}) satisfies the QND measurement
conditions~\cite{measure-1}. Since this two-mode Kerr interaction commutes
with the free Hamiltonians of both cavities, we can nondestructively measure
the photon number by observing the other cavity's conjugate observable.

\bigskip \bigskip \bigskip \bigskip

\section{\label{sec:five}CONCLUSION}

We have studied the Kerr nonlinearity in the vacuum induced by the radiation
pressure in typical optomechanical systems. Such nonlinear interaction can
be explicitly obtained via a BO approximation for an ideal case. Through a
squeezing effect, which should be experimentally observable via the
intensity spectrum, we verify the validity of the generalized BO
approximation by taking into account the dissipation and fluctuation of the
cavity field. Furthermore, we propose a two-mode Kerr interaction between
two cavity fields for quantum nondemolition measurements, which can be
realized by using two cavities sharing one oscillating mirror. It is
possible to nondestructively measure a cavity photon number by observing
another cavity's conjugate observable.\bigskip

\begin{acknowledgments}
We are grateful to Dr.~Lan Zhou for helpful discussions. C.P.~Sun was
supported by the NSFC with Grant Nos.90503003, NFRPC with Grant Nos.
2006CB921205 and 2005CB724508. FN was supported in part by the NSA, LPS,
ARO, and NSF Grant No. EIA-0130383.
\end{acknowledgments}

\end{document}